\documentclass[a4paper,11pt]{article}
\begin{document}

\author{A. Raiteri}
\title{A realistic interpretation of the density matrix\\
I: Basic concepts}
\date{December, 1998}
\maketitle

\section{Introduction}\label{sec_1}
The debate about the meaning of the wave function in quantum mechanics
is still going on, seventy years after the birth of the Copenhagen
interpretation \cite{copenh}. Despite the enormous successes achieved by
quantum mechanics and quantum field theory, we are not yet able to answer
the following simple question: should we consider the wave function as a
real objective field, or merely as a useful tool for computing probability
amplitudes?

There are (at least) two obstacles to a realistic interpretation of the wave
function:
\begin{enumerate}
\item For a N-body system, the wave function is defined in a 3N-dimensional
configuration space, while we would expect a ``real'' field to be defined
in ordinary 3-dimensional space. As a consequence, entanglement between
states of distant systems is allowed, and this gives rise to the
counter-intuitive behaviour known as ``quantum non-locality''.
\item The wave function spreads out with time. This property is strictly
related to the Heisenberg uncertainty principle: if a particle is well
localized in position space at a certain time, then it will be poorly
localized in momentum space, and therefore the dispersion of the wave packet
will grow with time. Our experience with macroscopic objects tells us that
they do not spread out with time, but in this case the quantum mechanical
explanation may be that for a macroscopic object the time needed to obtain
an observable dispersion of its wave packet is greater than the age of the
universe. However, microscopic objects too are always detected in small
space regions and it is hard to believe that immediately before their
detection they were spread out at long distances from the detection point.
\end{enumerate}

In this paper we conjecture that maybe the two above-mentioned problems
are a consequence of the particular choice of representation for the wave
function: with a suitable change of representation, it is possible to find
solutions to the wave equations which do not spread out with time. We'll see
that this result can be obtained at the cost of introducing a second
space-time coordinate $x_D$; we will consider this new coordinate as an
``auxiliary'' coordinate, in addition to the ``physical'' coordinate $x_S$.
Roughly speaking, the wave function will be spread out in the $x_D$
direction, while being perfectly localized in the ``physical'' $x_S$
direction.

We will first examine the non-relativistic Schr\"odinger equation (Section
\ref{sec_2}) and then the Dirac equation (Section \ref{sec_3}), both for
free fields and for interacting fields (we will focus on the electromagnetic
interaction); finally (Section \ref{sec_4}) we will draw some conclusions
and propose some possible interpretations of the results obtained in
Sections \ref{sec_2} and \ref{sec_3}.

\section{The non-relativistic Schr\"odinger equation}\label{sec_2}
Let's consider a quantum system describing a non-relativistic particle
moving in one dimension under the effect of an external potential
$V\!\left(x\right)$. If we choose the basis formed by the eigenvectors
$|x\rangle$ of the position operator Q, then the representation
of an arbitrary state is given by the wave function
$\psi\left(x\right)=\langle x|\psi\rangle$ and the Schr\"odinger equation is
\begin{equation}   \label{eq_1}
i\hbar\frac{\partial\psi}{\partial t}=\textrm{H}\,\psi=
-\frac{\hbar^2}{2m}\frac{\partial^2\psi}{\partial x^2}+V\!\left(x\right)\psi
\end{equation}
The time evolution of the system is a unitary transformation, and can be
written shortly as $\psi\left(x,t\right)=e^{-\frac{i}{\hbar}
\textrm{\scriptsize H}t}\psi\left(x,0\right)$. Besides, the position
operator Q and the momentum operator P are hermitian operators defined by
\begin{equation}\label{eq_2}
\textrm{Q}\,\psi=x\,\psi\qquad\qquad\qquad\quad
\textrm{P}\psi=-i\hbar\frac{\partial\psi}{\partial x}
\end{equation}
and they too generate unitary transformations of the form $e^{i\alpha
\textrm{\scriptsize Q}}$ and $e^{i\alpha\textrm{\scriptsize P}}$; the
transformations generated by the momentum operator are simply the space
displacements, i.e.\ $e^{i\,\Delta x\,\textrm{\scriptsize P}/\hbar}\:\psi
\left(x\right)=\psi\left(x+\Delta x\right)$.

We now define a new field $\varphi\left(x,y\right)$, which depends on the
wave function $\psi\left(x\right)$ through the following relation:
\begin{equation}\label{eq_3}
\varphi\left(x,y\right)=\psi\left(x\right)\psi^*\left(y\right)
\end{equation}
If we apply an arbitrary unitary transformation to the wave function $\psi
\left(x\right)$, the field $\varphi\left(x,y\right)$ will undergo a
corresponding transformation. Through this correspondence, we can obtain
the form of the generators in a new representation of the group of unitary
transformations. In the case of the hamiltonian operator H, which generates
the time evolution of the system, we easily obtain:
\begin{equation}   \label{eq_4}
i\hbar\frac{\partial\varphi}{\partial t}=\textrm{H}\,\varphi
=-\frac{\hbar^2}{2m}\left(\frac{\partial^2\varphi}{\partial x^2}-
\frac{\partial^2\varphi}{\partial y^2}\right)+\left[V\!\left(x\right)-
V\!\left(y\right)\right]\varphi
\end{equation}
Equation (\ref{eq_4}) is equal to the Schr\"odinger equation for density
matrices, which defines the time evolution for mixed states, and therefore
we will sometimes refer to the new representation as to the ``density matrix
representation''; in this paper, however, we will try to interpret the field
$\varphi$ as a real objective field, rather than just a tool for computing
probabilities. As for the generators Q and P, in the new representation we
obtain the expressions
\begin{equation}\label{eq_5}
\textrm{Q}\,\varphi=\left(x-y\right)\varphi\qquad\qquad\qquad\quad
\textrm{P}\varphi=-i\hbar\left(\frac{\partial\varphi}{\partial x}
+\frac{\partial\varphi}{\partial y}\right)\quad
\end{equation}

The definitions (\ref{eq_4}) and (\ref{eq_5}) can be used for an
arbitrary complex field $\varphi\left(x,y\right)$, not only for the fields
expressed by the relation (\ref{eq_3}). However, we will consider only
fields for which
\begin{equation}\label{eq_6}
\varphi\left(y,x\right)=\varphi^*\left(x,y\right)
\end{equation}
This condition is needed to obtain real values for the observable quantities,
as shown by expressions (\ref{eq_13})-(\ref{eq_15}) below.
From (\ref{eq_6}) it follows that we could use a real representation for
the field $\varphi$, simply by defining the real field
\begin{equation}\label{eq_7}
\varphi_R=Re\left\{\varphi\right\}+Im\left\{\varphi\right\}
\end{equation}
Indeed, $Re\left\{\varphi\right\}$ is symmetric and $Im\left\{\varphi
\right\}$ is antisymmetric with respect to the permutation $x\leftrightarrow
y$, and therefore they can be unambiguously obtained from the real field
$\varphi_R$. However, in the rest of this section we will use the complex
representation because it produces simpler expressions.

Now we want to define in the density matrix representation the observable
quantities, which are the position, the momentum and the energy. In the
original representation the natural definitions are the following:
\begin{eqnarray}
Q&=&\int\psi^*\left(x\right) x\ \psi\left(x\right)\textrm{d}x\label{eq_8}
\\[8pt] P&=&\int\psi^*\left(x\right)\left(-i\hbar\frac{\textrm{d}}
{\textrm{d}x}\right)\psi\left(x\right)\textrm{d}x\label{eq_9}\\[8pt]
E&=&\int\psi^*\left(x\right)\left(-\frac{\hbar^2}{2m}\frac{\textrm{d}^2}
{\textrm{d}x^2}+V\left(x\right)\right)\psi\left(x\right)\textrm{d}x
\label{eq_10}
\end{eqnarray}
In the probabilistic interpretation of (\ref{eq_1}), the above definitions
are simply the mean values of the corresponding quantum observables. However,
if we consider (\ref{eq_1}) as a classical field equation and set
$V\!\left(x\right)=0$, then it is easily seen that (\ref{eq_9}) is the
conserved quantity associated to the space invariance of the system,
(\ref{eq_10}) is the conserved quantity associated to the time invariance
of the system, while (\ref{eq_8}) is the center of mass, since
(\ref{eq_8}) and (\ref{eq_9}) satisfy the following classical relation
\begin{equation}\label{eq_11}
Q\left(t\right)=\frac{P}{m}\ t+Q\left(0\right)
\end{equation}
Besides, (\ref{eq_10}) is exactly the hamiltonian functional needed for
applying to (\ref{eq_1}) the canonical hamiltonian formalism. Thus the
definitions (\ref{eq_8}), (\ref{eq_9}) and (\ref{eq_10}) for position,
momentum and energy are reasonable also within a classical interpretation
of the Schr\"odinger equation (\ref{eq_1}).

For extending the above definitions to our new representation it is
convenient to introduce a new pair of space coordinates
\begin{equation}\label{eq_12}
x_S=\frac{1}{2}\left(x+y\right)\qquad\qquad\qquad x_D=y-x
\end{equation}
In the rest of this section we will write loosely $\varphi\left(x,y\right)$
or $\varphi\left(x_S,x_D\right)$, meaning the same field expressed in two
different coordinate systems. By means of the relation (\ref{eq_3}) we
obtain the following expressions in the new representation:
\begin{eqnarray}
Q&=&\int{x_S\ \varphi\left(x_S,x_D\right)\Big|_{x_D=0}\textrm{d}x_S}
\label{eq_13}\\[12pt]
P&=&i\hbar\int{\left.\frac{\displaystyle\partial\varphi\left(x_S,x_D\right)}
{\displaystyle\partial x_D}\right|_{x_D=0}\!\!\textrm{d}x_S}
\label{eq_14}\\[8pt]
E&=&\int{\left[-\frac{\displaystyle\hbar^2}{\displaystyle2m}\frac
{\displaystyle\partial^2\varphi\left(x_S,x_D\right)}{\displaystyle\partial
x_D^2}+V\!\left(x_S\right)\:\varphi\left(x_S,x_D\right)
\right]_{x_D=0}\!\!\textrm{d}x_S}\label{eq_15}
\end{eqnarray}
It can be shown that $P$ and $E$ are still conserved in the case
$V\!\left(x\right)=0$, while in the same case the classical relation
(\ref{eq_11}) is still satisfied.

Thus the physical quantities depend only on the value of the field $\varphi$
(and its space derivatives) in the region where $x_D=0$, or $x=y$. Therefore
we are induced to consider $x_S$ as the ``physical'' space coordinate, while
$x_D$ will be considered as an ``auxiliary'' coordinate, which has physical
effects only around the point $x_D=0$. Besides it can be shown that the
momentum operator P, as defined in (\ref{eq_5}), satisfies the following
relation:
\begin{equation}\label{eq_16}
e^{i\,\Delta x\,\textrm{\scriptsize P}/\hbar}\:\varphi\left(x_S,x_D\right)=
\varphi\left(x_S+\Delta x,x_D\right)
\end{equation}
Since we already know that the momentum operator generates the spatial
displacements, expression (\ref{eq_16}) adds more evidence to our belief
that $x_S$ is the ``physical'' position coordinate.

Now we want to express in the density matrix representation the unitary
transformation which allows to switch from the position space to the
momentum space. We know that this transformation for the original
Schr\"odinger equation is given by the Fourier transform
\begin{equation} \label{eq_17}
\Psi\left(k\right)=\frac{1}{\sqrt{2\pi}}\int{\psi\left(x\right)e^{-i k x}
\mathrm{d}x}
\end{equation}
We will again use the fundamental relation (\ref{eq_3}), exploiting the
property that it commutes with all unitary transformations, and then extend
the result to an arbitrary field $\varphi$. The final expression is
\begin{equation} \label{eq_18}
\Phi\left(k_x,k_y\right)=\frac{1}{2\pi}\int\!\int{\varphi\left(x,y\right)
e^{-i k_x\!x+i k_y\!y}\mathrm{d}x\,\mathrm{d}y}
\end{equation}
From (\ref{eq_18}) we easily find that in momentum space the operator P
is given by
\begin{equation}\label{eq_19}
\textrm{P}\,\Phi=\left(p_x-p_y\right)\,\Phi
\end{equation}
where we set $p_x=\hbar k_x$ and $p_y=\hbar k_y$; we note that expression
(\ref{eq_19}) has exactly the same form as the definition (\ref{eq_5}) for
the position operator Q. Besides, the quantity $P$ defined by (\ref{eq_14})
is expressed in momentum space by
\begin{equation}\label{eq_20}
P=\int{p_S\:\Phi\left(k_S,k_D\right)\Big|_{k_D=0}\textrm{d}k_S}
\end{equation}
where we have defined
\begin{equation}\label{eq_21}
k_S=\frac{1}{2}\left(k_x+k_y\right)\qquad\qquad\qquad k_D=k_y-k_x
\end{equation}
while obviously $p_S=\hbar\,k_S$. Again we note that expression (\ref{eq_20})
has exactly the same form as the definition (\ref{eq_13}) of the quantity
$Q$ in position space. This formal equivalence between the operators P and
Q, and between the observable quantities $P$ and $Q$, extends to our new
representation the simmetry between position space and momentum space which
is well known in non-relativistic quantum mechanics.

In the free particle case, where $V\!\left(x\right)\!=\!0$, the time
evolution of the system in momentum space is simply given by
\begin{equation}\label{eq_22}
i\hbar\frac{\partial\Phi}{\partial t}=
\left(\frac{p_x^2}{2m}-\frac{p_y^2}{2m}\right)\Phi
\end{equation}
while the energy $E$ becomes
\begin{equation}\label{eq_23}
E=\int{\frac{p_S^2}{2m}\:\Phi\left(k_S,k_D\right)\Big|_{k_D=0}
\textrm{d}k_S}
\end{equation}
From (\ref{eq_20}) and (\ref{eq_23}) we are again induced to consider $k_S$
as the ``physical'' momentum coordinate, while $k_D$ will be considered as
an ``auxiliary'' coordinate.

Now we will answer a very interesting question: what happens to the
Heisenberg uncertainty principle in the density matrix representation? In
the original representation the Heisenberg principle follows from the fact
that the operators Q and P do not commute:
\begin{equation}\label{eq_24}
\left[\,\textrm{Q}\:,\textrm{P}\,\right]=\textrm{Q}\:\textrm{P}
-\textrm{P}\:\textrm{Q}=i\hbar
\end{equation}
On the contrary, it can be easily shown that in our new representation the
operators P and Q, as defined by (\ref{eq_5}), do indeed commute. So it
seems that the Heisenberg principle is not true in this representation,
and indeed we will now show that there exist fields $\varphi$ which are
perfectly localized both in position space and in momentum space: let's
consider the field
\begin{equation}\label{eq_25}
\varphi_0\left(x_S,x_D\right)=\delta\left(x_S-x_0\right)e^{-i k_0 x_D}
\end{equation}
Having identified $x_S$ as the ``physical'' position, we can say that
$\varphi_0$ is perfectly localized at the position $x_0$, since we have
$\varphi_0=0$ for $x_S\ne x_0$. Applying the transformation (\ref{eq_18})
to $\varphi_0$ we obtain in momentum space
\begin{equation}\label{eq_26}
\Phi_0\left(k_S,k_D\right)=\delta\left(k_S-k_0\right)e^{i k_D x_0}
\end{equation}
and again, having identified $k_S$ as the ``physical'' momentum coordinate,
we can say that $\Phi_0$ is perfectly localized in momentum space, with
momentum $p_0=\hbar\,k_0$. Besides, applying to $\varphi_0$ the expression
(\ref{eq_13}), we indeed obtain $Q=x_0$, while applying to $\varphi_0$ the
expression (\ref{eq_14}) we obtain $P=\hbar\,k_0=p_0$ as expected; the same
value for $P$ is obtained in momentum space, i.e.\ applying the expression
\mbox{(\ref{eq_20}) to $\Phi_0$}.

In the case of a free particle, where $V\!\left(x\right)\!=\!0$, we easily
find that the energy, obtained from (\ref{eq_15}) or from (\ref{eq_23}), has
the correct classical value, i.e.\ $E=\hbar^2 k_0^2/2m=p_0^2/2m$. But there
is a far more interesting result: by setting
\begin{equation}\label{eq_27}
x_0=v_0 t\qquad\qquad\qquad\quad
k_0=\frac{p_0}{\hbar}=\frac{m v_0}{\hbar}\qquad\qquad
\end{equation}
which are the classical position and momentum for a particle moving with
constant speed $v_0$, we find that (\ref{eq_25}) is a solution of the
motion equations (\ref{eq_4}). Thus (\ref{eq_25}) and (\ref{eq_27})
define a field $\varphi_0\left(x_S,x_D,t\right)$ which can be interpreted
as a non-relativistic point-like particle moving at constant speed $v_0$
with position and momentum perfectly known at every time $t$. This result,
which seemed incompatible with quantum mechanics, has been obtained simply
by carrying the Schr\"odinger equation (\ref{eq_1}) to the density matrix
representation.

Now we turn to the general case of a potential $V\left(x\right)\ne0$. We
are interested in finding the stationary states, i.e.\ those states which
are sinusoidal functions of the time $t$. In the original representation,
the stationary states $\psi_\alpha$ were the eigenfunctions of the
hamiltonian
\begin{equation}\label{eq_28}
i\hbar\frac{\partial\psi_\alpha}{\partial t}=\textrm{H}\,\psi_\alpha=
E_\alpha\psi_\alpha
\end{equation}
while the set of all eigenvalues $E_\alpha$ was the energy spectrum of the
system. The same is true in our new representation, but now the hamiltonian
$H$ is expressed by equation (\ref{eq_4}). It is not difficult to show that
in our case the eigenfunctions $\varphi_{\alpha\beta}$ and the eigenvalues
$E_{\alpha\beta}$ are simply given by
\begin{equation}\label{eq_29}
\varphi_{\alpha\beta}\left(x,y\right)=\psi_\alpha\left(x\right)\psi_\beta^*
\left(y\right)\qquad\qquad\quad E_{\alpha\beta}=E_{\alpha}-E_{\beta}\quad
\end{equation}
Thus the natural frequencies in our representation correspond to the
difference between two energy levels of the original quantum system. But we
know very well that the individual energy levels are not really observable;
the only observable quantities are exactly the ``jumps'' between different
energy levels expressed by (\ref{eq_29}). Therefore we can say that we have
reproduced the same energy spectrum as the original quantum system; besides,
we have eliminated the ambiguity about the energy zero-point which is
typical of all quantum systems.

In conclusion, simply by changing the representation of the Schr\"odinger
equation we have succeeded in bringing together two concepts which are
widely held as incompatible: on one hand we describe point-like particles,
with position and momentum perfectly known at every time $t$, on the other
hand we allow the existence of discrete energy spectra. Besides, in the
case $V\!\left(x\right)\!=\!0$ we can add together an arbitrary number
of solutions of the form (\ref{eq_25}), thus obtaining a system of
non-interacting particles; the field describing this system depends always
on the same two variables $x_S$ and $x_D$, so the dimension of the space
on which the field is defined does not change with the number of the
particles. Finally, the transition to our new representation involves the
definition of an ``auxiliary'' position coordinate $x_D$, in addition to the
``physical'' coordinate $x_S$. The existence of extra space-time dimensions
is obviously not an original idea: it was first proposed by Kaluza and then
developed by Klein \cite{kalkle}. In recent years, I find very interesting
the attempt made by Hasselmann \cite{has} to derive all elementary
particles properties from Einstein's gravitational field equations in a
higher-dimensional matter-free space. Our model has in common with the
cited works the periodicity of the solutions with respect to the extra
space dimension, as shown by (\ref{eq_25}).

\section{The Dirac equation and the electromagnetic interaction}\label{sec_3}
We now try to extend the results obtained in Section \ref{sec_2} to the
case of a relativistic wave equation. We choose the Dirac equation, rather
than the Klein-Gordon equation, for the simple reason that all known
elementary particles are half-spin fermions, while no elementary spinless
particle is known to exist. We start from the non-covariant form of the
Dirac equation, which separates the time derivative from the space
derivatives:
\begin{equation}\label{eq_30}
i\,\frac{\partial\psi}{\partial t}=\left(-i\,\alpha_k\partial_k+m\,\beta
\right)\psi=\textrm{H}_0\,\psi
\end{equation}
where the implicit sum is over $k=1,2,3$.

The operator $\textrm{H}_0$ defined by equation (\ref{eq_30}) generates a
unitary transformation on the spinor field $\psi\left(\mathbf{x}\right)$,
where now $\psi$ depends upon a three-dimensional space coordinate
$\mathbf{x}$; we then follow the same method already seen in the previous
section and define a new representation of the group of unitary
transformations by means of the fundamental relation:
\begin{equation}\label{eq_31}
\varphi\left(\mathbf{x},\mathbf{y}\right)=\psi\left(\mathbf{x}\right)
\psi^\dag\left(\mathbf{y}\right)
\end{equation}
The time evolution in the new representation is then given by
\begin{equation}\label{eq_32}
i\,\frac{\partial\varphi}{\partial t}=\textrm{H}_0\,\varphi=
-i\left(\alpha_k\,\partial_{x_k}\varphi+\partial_{y_k}\varphi\,\alpha_k
\right)+m\left(\beta\,\varphi-\varphi\,\beta\right)
\end{equation}
We now extend equation (\ref{eq_32}) to an arbitrary $4\times4$ complex
matrix $\varphi\left(\mathbf{x},\mathbf{y}\right)$; however, as in the
previous section, to obtain real values for the  observable quantities the
anti-hermitian part of the matrix field $\varphi$ must vanish, and therefore
we will only consider fields $\varphi\left(\mathbf{x},\mathbf{y}
\right)$ satisfying the relation
\begin{equation}\label{eq_33}
\varphi\left(\mathbf{y},\mathbf{x}\right)=
\varphi^\dag\left(\mathbf{x},\mathbf{y}\right)
\end{equation}
Again we could use a real representation by defining $\varphi_R=Re\left\{
\varphi\right\}+Im\left\{\varphi\right\}$, but we prefer the complex
representation because it produces simpler expressions.

We are now able to find solutions to the motion equations (\ref{eq_32}) which
are perfectly localized in physical space and do not depend on time; these
solutions are thus good candidates to represent point-like particles at rest.
To obtain this result, we first define a new pair of space coordinates
\begin{equation}\label{eq_34}
\mathbf{x_S}=\frac{1}{2}\left(\mathbf{x}+\mathbf{y}\right)\qquad\qquad
\qquad\mathbf{x_D}=\mathbf{y}-\mathbf{x}\quad
\end{equation}
and then choose a particular representation for the matrices $\beta$ and
$\alpha_k$
\begin{equation}\label{eq_35}
\beta=\left[\begin{array}{rr}I&0\\[2pt]0&-I\end{array}\right]
\qquad\qquad\qquad\alpha_k=\left[\begin{array}{rr}0\ &\ \sigma_k\\[2pt]
\sigma_k&\ 0\ \end{array}\right]
\end{equation}
where $\sigma_k$ are the well known Pauli matrices, while $I$ is the
$2\times2$ identity matrix. Our solutions are then given by:
\begin{eqnarray}
\varphi_A&=&\frac{1}{4m}\left[
\begin{array}{cccc}
4\,m\,\delta\ &0\quad\ &i\,\partial_3\delta&
\ i\,\partial_1\delta+\partial_2\delta\\
0\ &0\quad\ &0&\ 0\\
-i\,\partial_3\delta\ &0\quad\ &0&\ 0\\
\partial_2\delta-i\,\partial_1\delta\ &0\quad\ &0&\ 0
\end{array}\right]\label{eq_36}\\[8pt]
\varphi_B&=&\frac{1}{4m}\left[
\begin{array}{cccc}
\ 0&0&0&0\\
\ 0&4\,m\,\delta&i\,\partial_1\delta-\partial_2\delta\ &
-i\,\partial_3\delta\\
\ 0&-i\,\partial_1\delta-\partial_2\delta&0&0\\
\ 0&i\,\partial_3\delta&0&0
\end{array}\right]\qquad\label{eq_37}\\[8pt]
\varphi_C&=&\frac{1}{4m}\left[
\begin{array}{cccc}
0\ &0&-i\,\partial_3\delta&\ 0\ \\
0\ &0&\partial_2\delta-i\,\partial_1\delta&\ 0\ \\
i\,\partial_3\delta\ &i\,\partial_1\delta+\partial_2\delta&-4\,m\,\delta&
\ 0\ \\
0\ &0&0&\ 0\
\end{array}\right]\label{eq_38}\\[8pt]
\varphi_D&=&\frac{1}{4m}\left[
\begin{array}{cccc}
0&\ 0&\quad 0\ &-i\,\partial_1\delta-\partial_2\delta\\
0&\ 0&\quad 0\ &i\,\partial_3\delta\\
0&\ 0&\quad 0\ &0\\
i\,\partial_1\delta-\partial_2\delta&\ -i\,\partial_3\delta&
\quad 0\ &-4\,m\,\delta
\end{array}\right]\label{eq_39}
\end{eqnarray}
where we wrote simply $\delta$ for $\delta\left(\mathbf{x_S}\right)$. If
we identify, as in the previous section, the physical position with the
coordinate $\mathbf{x_S}$, then the solutions (\ref{eq_36})-(\ref{eq_39})
are perfectly localized in the physical position $\mathbf{x_S}=0$. At this
point we would like to find solutions corresponding to particles moving at
constant speed; these solutions will be easily obtained from
(\ref{eq_36})-(\ref{eq_39}) by means of Lorentz transformations, but first
we must investigate the Lorentz covariance of our new representation.

There is only one way to recover a formal Lorentz covariance in our new
representation: since special relativity treats position and time as
components of the same vector $x^{\scriptstyle\mu}$, and since we have
duplicated the position coordinate $\mathbf{x}$, we inevitably must
duplicate also the time coordinate $t$. Thus we are led to introduce two
time coordinates $t_x$,$\,t_y$ and a second pair $t_S$,$\,t_D$ obtained
through a relation analogous to (\ref{eq_34}). In expressions (\ref{eq_31})
and (\ref{eq_33}) then the three-vectors $\mathbf{x}$ and $\mathbf{y}$ must
be replaced by the four-vectors $x\equiv\left(t_x,\mathbf{x}\right)$
and $y\equiv\left(t_y,\mathbf{y}\right)$, while the motion equations
(\ref{eq_32}) are extended to the following covariant form
\begin{equation}\label{eq_40}
i\,\gamma^\mu\frac{\partial}{\partial x^\mu}\,\varphi\!\left(x,y\right)-
m\,\varphi\!\left(x,y\right)=0
\end{equation}
which is formally identical to the covariant Dirac equation in the usual
representation. Equation (\ref{eq_40}) can be written in the
equivalent form
\begin{equation}\label{eq_41}
i\frac{\partial}{\partial y^\mu}\,\varphi\!\left(x,y\right)\gamma^0
\gamma^\mu+m\,\varphi\!\left(x,y\right)\gamma^0=0
\end{equation}
which is obtained by taking the adjoint of (\ref{eq_40}) and by exchanging
$x$ with $y$. If we separate the time derivative from the space derivatives,
(\ref{eq_40}) and (\ref{eq_41}) become
\begin{eqnarray}
i\,\frac{\partial\varphi}{\partial t_x}=
-i\,\alpha_k\,\partial_{x_k}\varphi+m\beta\,\varphi\label{eq_42}\\[4pt]
i\,\frac{\partial\varphi}{\partial t_y}=
-i\,\partial_{y_k}\varphi\,\alpha_k-m\,\varphi\,\beta\label{eq_43}
\end{eqnarray}
where we applied the usual definitions $\beta\!=\!\gamma^0$ and $\alpha_k\!=
\!\gamma^0\,\gamma^k$.
By adding together (\ref{eq_42}) and (\ref{eq_43}), we obtain the derivative
with respect to $t_S=\left(t_x+t_y\right)/2$, which is equal to the motion
equations (\ref{eq_32}) in the case $t_x=t_y=t$; by subtracting (\ref{eq_42})
from (\ref{eq_43}) and dividing by two, we obtain the derivative with respect
to $t_D\!=t_y\!-t_x$.

As for the Lorentz transformations, it is well known that an infinitesimal
Lorentz transformation acts on a four-vector $x^\mu$ and on a spinor $\psi$
in the following way:
\begin{eqnarray}
x'^\mu&=&x^\mu+{\epsilon^\mu}\!_\nu\,x^\nu\label{eq_44}\\\label{eq_45}
\psi'\,&=&\psi+\frac{1}{4}\,\epsilon_{\mu\nu}\gamma^\mu\gamma^\nu\psi\qquad
\end{eqnarray}
where $\epsilon_{\mu\nu}+\epsilon_{\nu\mu}=0$. By means of the fundamental
relation (\ref{eq_31}), we extend (\ref{eq_45}) to our new representation,
obtaining:
\begin{equation}\label{eq_46}
\varphi'\gamma^0=\varphi\,\gamma^0+
\frac{1}{4}\,\epsilon_{\mu\nu}\!\left(\gamma^\mu\gamma^\nu
\varphi\,\gamma^0+\varphi\,\gamma^0\gamma^\nu\gamma^\mu\right)
\end{equation}
We then separate the boosts from the spatial rotations rewriting
(\ref{eq_46}) as follows
\begin{equation}\label{47}
\varphi'=\varphi+i\,\mbox{\boldmath$\epsilon$}_\mathbf{R}\cdot\mathbf{R}\,
\varphi+i\,\mbox{\boldmath$\epsilon$}_\mathbf{B}\cdot\mathbf{B}\,\varphi
\end{equation}
where the vectors $\mbox{\boldmath$\epsilon$}_\mathbf{R}$ and
$\mbox{\boldmath$\epsilon$}_\mathbf{B}$ are
\begin{equation}\label{48}
\mbox{\boldmath$\epsilon$}_\mathbf{R}=\left[\begin{array}{c}\epsilon_{23}\\
\epsilon_{31}\\\epsilon_{12}\end{array}\right]\qquad\qquad\qquad
\mbox{\boldmath$\epsilon$}_\mathbf{B}=\left[\begin{array}{c}\epsilon_{01}\\
\epsilon_{02}\\\epsilon_{03}\end{array}\right]\qquad
\end{equation}
and the components of the two vector operators $\mathbf{R}$ and $\mathbf{B}$
are expressed by
\begin{eqnarray}
R_k\,\varphi&=&-\frac{1}{2}\left[\begin{array}{cc}\sigma_k&0\\0&\sigma_k
\end{array}\right]\varphi+\frac{1}{2}\,\varphi\left[\begin{array}{cc}
\sigma_k&0\\0&\sigma_k\end{array}\right]\label{eq_49}\\[12pt]
B_k\,\varphi&=&-\frac{i}{2}\,\alpha_k\,\varphi-\frac{i}{2}\,\varphi
\,\alpha_k\label{eq_50}
\end{eqnarray}
with $\alpha_k$ given by (\ref{eq_35}).

Now we are able to find solutions of the motion equations representing
particles moving with arbitrary speed $\mathbf{v}\!=\!v\,\mathbf{n}$. Let's
call $\varphi_0$ one of the four solutions (\ref{eq_36})-(\ref{eq_39})
representing particles at rest. The first step is to solve equations
(\ref{eq_42}) and (\ref{eq_43}) with respect to $t_D$, obtaining a
complete solution $\varphi_0\left(\mathbf{x_S},t_D\right)$ independent from
$\mathbf{x_D}$ and $t_S$; the solutions (\ref{eq_36})-(\ref{eq_39})
are recovered in the case $t_D=0$. Then we apply a boost with speed $v$
along the unit vector $\mathbf{n}$, which corresponds to the following
transformations:
\begin{eqnarray}
\varphi'&=&e^{i\,\xi\mathbf{n}\cdot\mathbf{B}}\,\varphi\\[6pt]
t'_S&=&t_S\cosh\xi+\left(\mathbf{x_S}\cdot\mathbf{n}\right)\sinh\xi\\[6pt]
\mathbf{x}'_\mathbf{S}&=&\mathbf{x_S}+\left(\mathbf{x_S}\cdot\mathbf{n}
\right)\mathbf{n}\left(\cosh\xi-1\right)+\mathbf{n}\,t_S\sinh\xi\qquad
\end{eqnarray}
where the parameter $\xi$ is defined by $\tanh\xi=v$; obviously $t_D$ and
$\mathbf{x_D}$ transform like $t_S$ and $\mathbf{x_S}$. Thus we obtain
the following solution:
\begin{equation}\label{eq_54}
\varphi'_\mathbf{v}\!\left(t'_S,\mathbf{x}'_\mathbf{S},t'_D,\mathbf{x}'_
\mathbf{D}\right)=e^{i\,\xi\mathbf{n}\cdot\mathbf{B}}\,\varphi_0\!
\left(\mathbf{x_S},t_D\right)
\end{equation}
It is clear that in the ``physical'' case $t'_D\!=\!\mathbf{x}'_\mathbf{D}\!
=\!0$ the solution (\ref{eq_54}) vanishes outside the region defined by
$\mathbf{x}'_\mathbf{S}\!=\!\mathbf{v}\,t'_S$; therefore it represents a
particle moving with speed $\mathbf{v}$ perfectly localized in the position
$\mathbf{x}'_\mathbf{S}\!=\!\mathbf{v}\,t'_S$.

Now we will express in our new representation the physical quantities
associated to the Dirac particles, which are energy, momentum, charge and
spin. Let's start from the charge $Q$, whose space density in the
original representation was the time component of the four-current
\begin{equation}\label{eq_55}
J^\mu\left(x\right)=\psi^\dag\left(x\right)\gamma^0\gamma^\mu
\psi\left(x\right)
\end{equation}
By means of the fundamental relation (\ref{eq_31}) we obtain in the new
representation the following expression
\begin{equation}\label{eq_56}
J^\mu\left(x\right)=\textrm{Tr}\left\{\,\varphi\left(x_S,x_D\right)
\gamma^0\gamma^\mu\right\}\bigg|_{x_S=x\,,\,x_D=0}
\end{equation}
The time conservation of the charge $Q$ is a consequence of the continuity
equation $\partial_\mu J^\mu\!=\!0$, which is true in both representations.
The explicit form of the charge $Q$ in the new representation is simply
given by
\begin{equation}\label{eq_57}
Q=\int\textrm{d}^3\!x\sum_{k=1}^4\varphi_{kk}\left(\mathbf{x_S},
\mathbf{x_D}\right)\bigg|_{\mathbf{x_S}=\mathbf{x}\,,\,\mathbf{x_D}=0}
\end{equation}

In the same way, starting from the energy-momentum tensor in the original
representation, we easily obtain the following expression for the same
tensor in our new representation:
\begin{equation}\label{eq_58}
{T\,^\mu\!}_\nu\left(x\right)=\textrm{Tr}\left\{-i\frac{\partial}
{\partial x_D^\nu}\,\varphi\left(x_S,x_D\right)\gamma^0\gamma^\mu\right\}
\Bigg|_{x_S=x\,,\,x_D=0}
\end{equation}
In both representations the conservation of energy and momentum is expressed
by the continuity equation $\partial_\mu T^{\,\mu\nu}\!=0$, while the
definitions of energy and momentum in the new representation are
\begin{equation}\label{eq_59}
E=\int\!\mathcal{E}\!\left(\mathbf{x}\right)\textrm{d}^3\!x\qquad\qquad
\qquad P_i=\int\!\mathcal{P}_i\!\left(\mathbf{x}\right)\textrm{d}^3\!x
\end{equation}
with the spatial densities $\mathcal{E}$ and $\mathcal{P}_i$ given by
\begin{eqnarray}
\mathcal{E}&=&T^{\,00}=\textrm{Tr}\left\{i\,\alpha_k\,\frac{\partial\,
\varphi}{\partial x_{D,k}}+m\,\beta\,\varphi\right\}\Bigg|_{\mathbf{x_S}=
\mathbf{x}\,,\,\mathbf{x_D}=0}\label{eq_60}\\[8pt]
\mathcal{P}_i&=&T^{\,0i}=\sum_{k=1}^4 i\,\frac{\partial\,\varphi_{kk}}
{\partial x_{D,i}}\,\bigg|_{\mathbf{x_S}=\mathbf{x}\,,\,\mathbf{x_D}=0}
\label{eq_61}
\end{eqnarray}

Finally, from the angular momentum tensor in the original representation,
we obtain the following expression for the same tensor in the new
representation:
\begin{eqnarray}
M^{\,\mu\lambda\nu}\left(x\right)&=&x^\lambda\,T^{\,\mu\nu}\left(x\right)
-x^\nu\,T^{\,\mu\lambda}\left(x\right)\nonumber\\[6pt]
&+&\textrm{Tr}\left\{\frac{i}{4}\,\varphi\left(x_S,x_D\right)
\gamma^0\gamma^\mu\left(\gamma^\lambda\gamma^\nu-\gamma^\nu\gamma^\lambda
\right)\right\}\bigg|_{x_S=x\,,\,x_D=0}
\label{eq_62}
\end{eqnarray}
In both representations the angular momentum conservation is expressed by
the continuity equation $\partial_\mu M^{\,\mu\lambda\nu}\!=0$. Besides
the angular momentum spatial densities are given by:
\begin{equation}\label{eq_63}
M^{0ij}\!\left(x\right)=\epsilon_{ijk}\Big(\mathcal{L}_k\!\left(
x\right)+\mathcal{S}_k\!\left(x\right)\Big)\qquad\qquad
i,j,k=1,2,3
\end{equation}
where $\epsilon_{ijk}$ is the rank three antisymmetric tensor; $\mathcal{L}$
is the orbital angular momentum density, while $\mathcal{S}$ is the spin
density. The spin components are then given by:
\begin{equation}\label{eq_64}
S_i=\int\!\mathcal{S}_i\!\left(\mathbf{x}\right)
\textrm{d}^3\!x=\int\!\textrm{d}^3\!x\ \textrm{Tr}\left\{\frac{1}{2}\left[
\begin{array}{cc}\sigma_i&0\\0&\sigma_i\end{array}\right]\varphi\left(
\mathbf{x_S},\mathbf{x_D}\right)\right\}
\Bigg|_{\mathbf{x_S}=\mathbf{x}\,,\,\mathbf{x_D}=0}
\end{equation}
From expressions (\ref{eq_56}), (\ref{eq_58}) and (\ref{eq_62}), it is clear
that all observable quantities depend on the value of the field $\varphi$
(and its space-time derivatives) in the region where $x_D=0$, or $x=y$.
Once again we are induced to consider $x_S$ as the ``physical'' space-time
coordinate, while $x_D$ is an ``auxiliary'' coordinate, which has observable
effects only around the point $x_D=0$.

We are now able to associate to the solutions defined in
(\ref{eq_36})-(\ref{eq_39}) their physical properties. From the
above definitions, we see that $\varphi_A$, $\varphi_B$,
$\varphi_C$ and $\varphi_D$ have energy $E\!=\!m$, momentum $\mathbf{P}\!=
\!0$ and spin components $S_1\!=\!S_2\!=\!0$. As for the charge $Q$ and the
third spin component $S_3$, we have the following situation: $\varphi_A$
has $Q\!=\!+1$ and $S_3\!=\!+1/2$, thus representing a spin up positron;
$\varphi_B$ has $Q\!=\!+1$ and $S_3\!=\!-1/2$, thus representing a spin down
positron; $\varphi_C$ has $Q\!=\!-1$ and $S_3\!=\!-1/2$, thus representing a
spin down electron; finally, $\varphi_D$ has $Q\!=\!-1$ and $S_3\!=\!+1/2$,
thus representing a spin up electron. Besides, if we apply a rotation or a
boost to these solutions, we are certain that the physical quantities
associated to the new solutions will have the usual relativistic values;
this is true because we have expressed all the observable quantities in a
covariant form.

Until now, we have shown that our new representation admits solutions
which can be interpreted as relativistic point-like particles moving at
constant speed in an arbitrary direction, with the correct values for
energy and momentum. In their rest frame, the particles have spin $S=1/2$,
while the spin orientation is arbitrary; as for the charge $Q$, we have
solutions with $Q=+1$, to be interpreted as positrons, and solutions with
$Q=-1$, to be interpreted as electrons. By adding together an arbitrary
number of such solutions, we obtain a system of non-interacting particles.
Although attractive, this description is far from complete: indeed, there
are a lot of solutions for which such particle interpretation is not
possible and our model does not explain why the physical solutions should
be acceptable while the non-physical solutions should be discarded.

At this point, we must put forward the following consideration: real
particles interact with each other and we can observe them only because
they interact with our measuring devices; therefore, all mathematical models
describing non-interacting particles are equally useless for explaining
physical phenomena. To complete our model, then, we need an interaction
between particles; we will choose the electromagnetic interaction, which has
provided the first striking application of quantum field theory, as well as
the first example of a modern gauge theory.

To introduce the electromagnetic interaction in our model, we return to the
original representation, where we make the usual substitution
\begin{equation}\label{eq_65}
\frac{\partial}{\partial x^\mu}\quad\to\quad\frac{\partial}{\partial x^\mu}
-i e A_\mu
\end{equation}
If we insert (\ref{eq_65}) in the non-covariant Dirac equation (\ref{eq_30}),
and then switch to our new representation by means of (\ref{eq_31}), we
obtain the following time evolution:
\begin{equation}\label{eq_66}
i\,\frac{\partial\varphi}{\partial t}=\textrm{H}_0\,\varphi
+e\,\textrm{H}_1\,\varphi
\end{equation}
where $\textrm{H}_0$ is still defined as in (\ref{eq_32}), while
$\textrm{H}_1$ is given by
\begin{equation}\label{eq_67}
\textrm{H}_1\,\varphi=A^0\!\left(\mathbf{y}\right)\varphi-A^0\!\left(
\mathbf{x}\right)\varphi+A^k\!\left(\mathbf{x}\right)\alpha_k\,\varphi
-A^k\!\left(\mathbf{y}\right)\varphi\,\alpha_k
\end{equation}
where, as usual, the implicit sum is over $k=1,2,3$.

As for the covariant equations (\ref{eq_40}) and (\ref{eq_41}), in the
presence of an external electromagnetic field they become:
\begin{eqnarray}
&i&\gamma^\mu\left(\frac{\partial}{\partial x^\mu}-i\,e\,A_\mu
\!\left(x\right)\right)\varphi-m\,\varphi=0
\label{eq_68}\\[6pt]
&i&\left(\frac{\partial}{\partial y^\mu}+i\,e\,A_\mu\!\left(y\right)
\right)\varphi\,\gamma^0\gamma^\mu+m\,\varphi\,\gamma^0=0\label{eq_69}
\end{eqnarray}
As in the free field case, (\ref{eq_69}) can be obtained by taking the
adjoint of (\ref{eq_68}) and by exchanging $x$ with $y$. The non-covariant
time evolution (\ref{eq_66}) can be recovered from (\ref{eq_68}) and
(\ref{eq_69}) simply by setting $t_x=t_y=t$.

Let's spend a few words about charge conjugation and gauge invariance. In
the original representation, if $\psi$ is a solution of the Dirac equation
interacting with an electromagnetic field
\begin{equation}\label{eq_70}
i\,\gamma^\mu\left(\partial_\mu-i\,e\,A_\mu\right)\psi-m\,\psi=0
\end{equation}
then the field
\begin{equation}\label{eq_71}
\psi'\!\left(x\right)=\gamma^2\psi^*\!\left(x\right)
\end{equation}
is a solution of the equation obtained from (\ref{eq_70}) by changing the
sign of the interaction parameter $e$:
\begin{equation}\label{eq_72}
i\,\gamma^\mu\left(\partial_\mu+i\,e\,A_\mu\right)\psi'-m\,\psi'=0
\end{equation}
The transformation (\ref{eq_71}) is called charge conjugation and depends
on the particular representation (\ref{eq_35}) we have chosen for the Dirac
matrices. It is easily seen that $\psi$ and $\psi'$ have the same charge
$Q\!=\!\int\psi^\dag\left(\mathbf{x}\right)\psi\left(\mathbf{x}\right)
\textrm{d}^3\!x\,$; we would expect the charge conjugation to change the
sign of the charge, as it happens for instance in the case of the
Klein-Gordon equation, but this is of course impossible for the Dirac
equation, since the charge is always positive in the original representation.
If we now switch to our new representation, we find that the transformation
(\ref{eq_71}) becomes
\begin{equation}\label{eq_73}
\varphi'\!\left(x,y\right)=-\gamma^2\varphi^T\!\left(y,x\right)\gamma^2
\end{equation}
Again we find that $\varphi'$ satisfies the motion equations obtained by
changing the sign of the interaction parameter $e$, and again $\varphi$ and
$\varphi'$ have the same charge, i.e.\ $Q'\!=\!Q$. However, if we now change
the sign of (\ref{eq_73}), we still have a solution of the same motion
equations, but now we have $Q'\!=\!-Q$. Therefore in our new
representation the correct definition for the charge conjugation is
\begin{equation}\label{eq_74}
\varphi'\!\left(x,y\right)=+\gamma^2\varphi^T\!\left(y,x\right)\gamma^2
\end{equation}
It can be shown that (\ref{eq_74}) changes the sign of the whole
four-current $J^\mu$, so that the field $\varphi'$ satisfies the conjugate
Maxwell equations for the electromagnetic field; this property was obviously
not true in the original representation. Besides, the transformation
(\ref{eq_74}) leaves unchanged all other observable quantities, namely
energy, momentum and spin, and this is exactly what we expect from a well
defined charge conjugation transformation; for instance, if we consider the
four solutions (\ref{eq_36})-(\ref{eq_39}), we easily see that $\varphi_D$
($\varphi_C$) is obtained by charge conjugation from $\varphi_A$
($\varphi_B$) and viceversa.

Turning now to the gauge invariance, we remember that in the original
representation a gauge transformation is given by a local phase rotation
\mbox{on the field $\psi$}
\begin{equation}\label{eq_75}
\psi'\left(x\right)=e^{i\,\theta\left(x\right)}\psi\left(x\right)
\end{equation}
together with the transformation
\begin{equation}\label{eq_76}
A'_\mu\left(x\right)=A_\mu\left(x\right)+\frac{1}{e}
\frac{\partial}{\partial x^\mu}\,\theta\left(x\right)
\end{equation}
for the electromagnetic field. Gauge invariance means that if $\psi$ and
$A^\mu$ are solutions of the motion equations, the same is true for $\psi'$
and $A'^\mu$. In our new representation (\ref{eq_75}) becomes
\begin{equation}\label{eq_77}
\varphi'\left(x,y\right)=e^{i\,\left[\theta\left(x\right)-\theta\left(y
\right)\right]}\varphi\left(x,y\right)
\end{equation}
and it is easily seen that the motion equations (\ref{eq_68}), or
(\ref{eq_69}), are invariant with respect to the transformations
(\ref{eq_76}) and (\ref{eq_77}), i.e.\ they are gauge invariant.

Until now we have considered the electromagnetic field as an external source,
rather than a dynamical component of our model. If we want to include the
field $A^\mu$ in our model, we have to define its time evolution
in interaction with the ``material'' field $\varphi$. The dynamical
behaviour of the electromagnetic field will be described by the classical
Maxwell equations:
\begin{equation}\label{eq_78}
\partial^\mu\left(\partial_\mu A_\nu-\partial_\nu A_\mu\right)=
\partial^\mu F_{\mu\nu}=J_\nu
\end{equation}
We simply have to define the current $J_\nu$ as a function of the field
$\varphi$. To obtain the electric current density from (\ref{eq_56})
we multiply by the unit electric charge $e$, obtaining:
\begin{equation}\label{eq_79}
J^\mu\left(x\right)=e\,\textrm{Tr}\left\{\,\varphi\left(x_S,x_D\right)
\gamma^0\gamma^\mu\right\}\bigg|_{x_S=x\,,\,x_D=0}
\end{equation}
The Maxwell equations (\ref{eq_78}) are again gauge invariant: the left-hand
side depends only on the gauge invariant tensor $F_{\mu\nu}$, while the
right-hand side depends only on the values of $\varphi$ for $x_D=0$, or
$x=y$, which are not affected by the \mbox{transformation (\ref{eq_77})}.

In conclusion, we have built a model which describes the ``material'' Dirac
field $\varphi$ together with the electromagnetic field $A^\mu$. The model
is defined by the time evolution of the field $\varphi$, (\ref{eq_68}) or
(\ref{eq_69}), and by the Maxwell equations (\ref{eq_78}) where the current
is given by (\ref{eq_79}). The field $\varphi$ depends on two space-time
coordinates $x_S$ and $x_D$; however, we point out that the duplication of
the time coordinate is needed only for formal requirements: the motion
equations (\ref{eq_32}) or (\ref{eq_66}) can be solved with respect to the
``physical'' time ($t_S=t$ and $t_D=0$) without even mentioning a second
time coordinate. On the contrary, the electromagnetic field $A^\mu$ depends
on a single space-time coordinate $x$ and its time evolution is given by
the classical Maxwell equations; it is interesting here to note that some
authors (see Marshall and Santos \cite{myth}) reject the quantum concept
of an electromagnetic field made up by point-like massless ``photons'', and
believe that all experimental data involving light fields can be explained
by means of the unquantized Maxwell equations.

\section{Discussion}\label{sec_4}
Let me explain the basic ideas (or maybe prejudices) on which the present
paper is built. The main idea is that quantum mechanics cannot be a
fundamental theory: I cannot force myself to believe that probabilities
are an objective feature of the physical world, and therefore I maintain
the old-fashioned belief that the fundamental laws of nature must be
deterministic and quantum mechanics must be just a statistical formulation
of these fundamental laws. The obvious example for clarifying this concept
is the classical motion of a free non-relativistic particle, with position
$x$ and momentum $p\,$: the ``wave equation''
\begin{equation}\label{eq_80}
\frac{\partial f}{\partial t}=-\frac{p}{m}\frac{\partial f}{\partial x}
\end{equation}
for the joint probability distribution $f\left(x,p\right)$ is just a
statistical formulation of the deterministic state equations
\begin{equation}\label{eq_81}
\dot{x}=\frac{p}{m}\qquad\qquad\qquad\qquad\dot{p}=0\qquad\qquad
\end{equation}
We are not worried about the ``collapse of the wave function'' at the time
when the position $x$ is measured, because we do not interpret $f\left(x,p
\right)$ as a real objective field; in any case, such interpretation would
be strongly hampered by the fact that $f\left(x,p\right)$ is defined in
configuration space, whose dimension depends on the number of particles
described.

Now, what mathematical form should have this deterministic law which will
finally replace quantum mechanics and quantum field theory? I am convinced
that it should be a classical (unquantized) field equation, satisfying the
principle of relativistic covariance. However, since until now all efforts
at finding this unified field equation have failed, we probably need to
extend the classical concept of field, to gain some room for maneuver.
In my opinion, the most natural extension is to allow for extra space-time
dimensions, following the idea which was originally proposed by Kaluza and
Klein \cite{kalkle}, and has been recently revived by Hasselmann \cite{has}.
Another possible choice could be to abandon the concept of a continuous
space-time, and suppose that the ultimate structure of space-time is
discrete (see for instance Budnik \cite{budnik}); however, besides having a
prejudice against ``objective probabilities'', I also have a prejudice in
favour of Newton's ``Natura non facit saltus'', basing this prejudice mainly
on the wish to maintain the beautiful space-time simmetries and conservation
laws which are strictly connected to the continuous structure of space-time.

Thus, we are looking for a relativistic field equation (or a system of
equations) defined on a space-time with more than four dimensions.
But what should be our starting point for this long and perilous search?
The authors cited above stated immediately their ambitious goal: they wanted
to unify gravity with all other known interactions (in the Kaluza-Klein case
it was just the electromagnetic interaction, while Hasselmann's ``metron''
model includes all Standard Model's interactions). Even if this is indeed
the final goal, I propose here a more cautious approach: instead of trying
to guess from the beginning the final form of our field equations, let's
start from the field equations on which quantum mechanics and quantum field
theory are based (the non-relativistic Schr\"odinger equation and the
relativistic Dirac equation) and let's try to extend them in such a way that
their solutions may now be interpreted as real objective fields.

In the introduction to this paper (Section \ref{sec_1}) I listed the two
main obstacles to a realistic interpretation of the wave function in
non-relativistic quantum mechanics: (1) the wave function spreads out with
time and (2) it is defined on a variable dimension configuration space.
Well, in Section \ref{sec_2} we saw that if we choose the Schr\"odinger
equation for density matrices, instead of the usual equation for pure
states, it is possible to find solutions which are perfectly localized both
in position space and in momentum space. In the free field case, these
solutions behave exactly as non-relativistic point-like particles moving at
constant speed, with the correct values for the observable quantities
(position, energy and momentum). By adding together an arbitrary number
of such solutions we obtain a field describing a system of non-interacting
particles; this field depends always on the same two space coordinates, the
``physical'' coordinate $x_S$ and the ``auxiliary'' coordinate $x_D$. Both
problems (1) and (2) are solved, and I believe that this is a first
(small) step in the right direction.

It is interesting here to note that some authors already consider the
Schr\"odinger equation for density matrices more fundamental than the
one for pure states. For instance Olavo \cite{olavo} explicitly names
the equation for density matrices ``First Schr\"odinger's Equation'',
while the equation for pure states, named ``Second Schr\"odinger's
Equation'', is obtained from the first in the less general case where
the density matrix is a product of pure states. Olavo also defines
the position operator and the momentum operator for density matrices and
shows that they commute, similarly to what we did in Section \ref{sec_2};
from this property Olavo infers the ``negation of ontological origin of
Heisenberg's uncertainty relations'' and I do indeed agree with him.
However, Olavo's purpose is to demonstrate that quantum mechanics can be
derived from classical mechanics, and therefore he is not interested
in giving a realistic interpretation of the density matrix.

A second example can be found in Mermin's ``Ithaca interpretation''
\cite{merm}; in this paper the author lists six desiderata for a
satisfactory interpretation of quantum mechanics, and then writes
``If you take Desideratum (5) seriously (i.e. generalized Einstein
locality), then there can be no more objective reality to the possible
different realizations of a density matrix, than there is to the different
possible ways of expanding a pure state in terms of different complete
orthonormal sets \dots\ In the case of an individual system the density
matrix must be a fundamental and irreducible objective property, whether
or not it is a pure state''. This last statement sounds quite similar to
the idea that I am suggesting in the present paper, i.e. the density matrix
as a real objective field. However, this resemblance disappears when we read
Desideratum (6), which states that probabilities are objective intrinsic
properties of individual physical systems; unlike Mermin, I still believe
that probability is ``just a way of dealing sistematically with our own
ignorance'', and not a fundamental feature of the physical world.

Let's turn now to the relativistic Dirac equation. We immediately see a new
obstacle to a realistic interpretation of its solutions: (3) the charge
density $\psi^\dag\psi$ is always positive, and therefore the classical
Dirac field does not allow us to describe particles with both positive and
negative charge, i.e.\ positrons and electrons. The second quantization
procedure solves point (3) by means of the anti-commuting properties of the
creation-annihilation operators, but the points (1) and (2) listed above
remain unsolved; furthermore, the final wave function (or state vector)
is defined in a Hilbert space whose dimension is dramatically increased.
On the contrary, in Section \ref{sec_3} we saw that, by extending to the
Dirac equation the same representation already used for the non-relativistic
Schr\"odinger equation, it is possible to find solutions which behave as
relativistic point-like particles moving at constant speed, with the correct
values for all observable quantities. By adding together an arbitrary number
of such solutions we can describe a system of non-interacting particles, and
the resulting field depends always on the same six space coordinates, three
``physical'' coordinates $\mathbf{x_S}$ and three ``auxiliary'' coordinates
$\mathbf{x_D}$. By introducing also two time coordinates (a ``physical''
time coordinate $t_S$ and an ``auxiliary'' time coordinate $t_D$) we can
recover the formal Lorentz covariance of the Dirac equation. Finally, in
the new representation the charge can have both signs, and more generally
it is possible to define a charge conjugation transformation which changes
the sign of the charge and leaves unchanged all other physical quantities.
Therefore, points (1), (2) and (3) are all solved, and I think that this is
a second step in the right direction.

Thus, for both the non-relativistic Schr\"odinger equation and the
relativistic Dirac equation we were able to find solutions which can be
interpreted as non-interacting point-like particles. At this point, however,
we must face a big problem: these solutions are only a limited subset of all
possible solutions, while for most solutions this particle interpretation
is not possible; until now our model does not explain why the ``particle''
solutions should be accepted while the other solutions should be discarded.
I will discuss this problem only for the Dirac equation, neglecting the
non-relativistic case for the following reasons: first, not all potentials
$V\left(x\right)$ have physical meaning; second, since non-relativistic
quantum mechanics is just an approximation for relativistic quantum field
theory, if we find an acceptable model for quantum field theory then the
corresponding model for quantum mechanics will follow in the
non-relativistic limit.

In my opinion, there is only one way to single out the ``particle''
solutions of the Dirac equation from all other solutions: we must add a
non-linear interaction to the model. The interaction must be non-linear for
obvious reasons: for instance, the field obtained by changing the sign to
a positive energy solution cannot be itself a solution, because it would
have negative energy (negative mass in its rest frame) and therefore it
would be non-physical. Thus, in the second part of Section \ref{sec_3}
we inserted in our model the electromagnetic interaction, by means of the
usual ``minimal coupling'' prescription. We also included in our model
the dynamical evolution of the electromagnetic field, in the form of the
classical Maxwell equations; the four-current in the Maxwell equations
depends on the Dirac field through equation (\ref{eq_79}). Our final model
provides a complete description of the mutual interactions between the
``material'' Dirac field and the ``gauge'' field, i.e.\ the classical
electromagnetic field.

Now the question is: does this final model really solve the problem of
eliminating the non-physical solutions? This will be true if the only stable
solutions for the ``material'' field $\varphi$ are the vacuum (i.e.\ the
solution $\varphi\!=\!0$) and the single particle solutions, which must be
strongly localized in physical space and must provide the correct values
for all observable quantities; we can relax this last requirement by
accepting some form of renormalization, i.e.\ the physical mass and charge
$m_{ph}$ and $e_{ph}$ could be different from the ``bare'' quantities $m$
and $e$ which appear in the model equations. Unfortunately, the task of
finding the solutions to our model (either by symbolic calculations or by
computer simulations) seems to me extraordinarily complex, so our question
will find no answer in the present paper. However, let's suppose that the
answer may be affirmative; if this were true, then we could
attempt to interpret our model as a deterministic alternative to quantum
electrodynamics. To reach this ambitious goal, we would have to satisfy
(at least) two further requirements: first, in scattering processes
our model should be able to reproduce the well known phenomenology of QED,
with the same probability amplitudes; second, by adding an external source
to our model (for instance the charge of the hydrogen atom nucleus), the
stable solutions should reproduce the same energy spectra which can be
obtained from standard QED.

On the contrary, if the answer to our question were a plain no, this
would not necessarily entail that all our previous efforts have been
useless; we could try to make some changes in our model, along one of
the following lines:
\begin{itemize}
\item Maybe the electromagnetic interaction alone is not enough to single
out the ``particle'' solutions from all other solutions: maybe it is
necessary to insert in our model the other known interactions, or even some
new (yet unknown) interaction, which would be responsible for the strong
localization of the electron and positron fields. My guess that the
electromagnetic interaction may suffice is based only on the fact that QED
can exist as an independent theory even if we know that it is really a part
of a more general theory, i.e.\ the Weinberg-Salam model for electroweak
interactions.
\item In this paper we suppose that the ``auxiliary'' space-time, indexed by
the coordinate $x_D$, is infinite and flat. It could as well be finite and
strongly curved, similarly to the fifth dimension in the Klein version of
the original Kaluza theory. If the curvature radius were really small
compared to macroscopic space-time distances, this could explain why we do
not observe the extra space-time dimensions.
\item Finally, we could look for a way to extend the electromagnetic field
to the extra space-time dimensions, obtaining a new field $A^\mu\left(x_S,
x_D\right)$ which would be equal to the classical electromagnetic field in
the case $x_D=0$. The resulting model should have a better simmetry between
the material field and the gauge field.
\end{itemize}
Of course, I have no absolute certainty that the method presented in this
paper will finally produce acceptable explanations of physical phenomena: I
just tried to find a new approach to deal with a very old problem. However,
it can be shown that if we apply this method to simple non-physical systems
(two- and three-dimensional Hilbert spaces) we indeed obtain deterministic
models equivalent to the corresponding quantum systems; in this case, the
density matrix representation becomes simply the adjoint representation of
the SU(2) and SU(3) groups.

In conclusion, I will point out some features of our model which could be
related to the debate about quantum non-locality; let's briefly summarize
the main results of this long lasting debate. In 1964 Bell \cite{bell}
showed that no local hidden variable (LHV) model can reproduce the same
probability correlations obtained from quantum mechanics; he did so by
constructing an explicit relation (the first famous Bell inequality) which
must be satisfied by every LHV model and which is violated by some quantum
states. In the early 80's, experiments were performed \cite{aspect1,aspect2}
which seemed to indicate that Bell inequalities are indeed violated by
nature. If this were true, then we would be forced to choose between
determinism and locality; for instance, Mermin \cite{merm} chooses to
abandon determinism and to retain locality. However, the debate about the
experimental violation of Bell inequalities is still open: it seems that
this violation can be inferred from the existing experimental data only by
introducing some supplementary assumptions (no enhancement, faithful
sampling, \dots) and therefore many authors \cite{kwiat,santos1} believe
that until now no incontrovertible violation has been really observed.
Besides, Marshall and Santos \cite{myth,santos2} are convinced that the
corpuscular interpretation of light plays an important role in the
above-mentioned assumptions, and were able to build a local ``wave'' theory
of light, derived from Maxwell's classical equations, whose predictions are
in good agreement with the experimental data \cite{marsh}.

The model presented in Section \ref{sec_3} of the present paper is expressed
in a relativistic covariant form and therefore it must satisfy the principle
of relativistic causality (space-like separated events cannot influence each
other and no interaction can propagate at superluminal speed). However, the
following features seem to indicate some form of non-local behaviour:
\begin{itemize}
\item Expression (\ref{eq_33}) establishes an instant relation between the
fields $\varphi\left(\mathbf{x},\mathbf{y}\right)$ and $\varphi\left(
\mathbf{y},\mathbf{x}\right)$. However, this non-local relation is confined
to extra-space, while in the physical case $\mathbf{x}=\mathbf{y}$ we obtain
a local hermiticity property for the matrix field $\varphi\left(\mathbf{x},
\mathbf{x}\right)$.
\item The solutions (\ref{eq_36})-(\ref{eq_39}) representing particles at
rest are perfectly localized in the physical position $\mathbf{x_S}\!=\!0$
and do not depend on the auxiliary coordinate $\mathbf{x_D}$. However, it
can be shown that the solutions representing moving particles, while being
perfectly localized at $\mathbf{x_S}\!=\!\mathbf{v}\,t$ when $\mathbf{x_D}\!
=\!0$, spread out along the $\mathbf{x_S}$ axis when $\mathbf{x_D}\!\ne\!0$.
Indeed, if $\mathbf{n}$ is the direction of motion, then the corresponding
field is different from zero inside the region defined by $|\mathbf{x_S}\!
\cdot\!\mathbf{n}-v\,t|\!\le\!v\,|\mathbf{x_d}\!\cdot\!\mathbf{n}|/2$.
Therefore, if we have two such solutions localized at two distant physical
points, their fields will anyway overlap in some extra-space regions. This
does not mean that the two fields interact, since in the free field case
the motion equations are linear, but in any case this is again a form of
non-locality in extra-space.
\item Finally, equation (\ref{eq_67}) clearly shows that the electromagnetic
field $A^\mu\left(\mathbf{z}\right)$ acts instantly on the field $\varphi
\left(\mathbf{x},\mathbf{y}\right)$ at all those points having $\mathbf{x}
\!=\!\mathbf{z}$ or $\mathbf{y}\!=\!\mathbf{z}$. Therefore an interaction
can propagate between two physical points $A=(\mathbf{z_A},\mathbf{z_A})$
and $B=(\mathbf{z_B},\mathbf{z_B})$ through the following mechanism: the
field $\varphi(\mathbf{z_A},\mathbf{z_A})$ acts instantly on the
electromagnetic field $A^\mu\left(\mathbf{z_A}\right)$ through the current
(\ref{eq_79}); the field $A^\mu\left(\mathbf{z_A}\right)$ acts, again
instantly, on a extra-space point $C=(\mathbf{z_A},\mathbf{z_B})$; finally,
the interaction propagates from $C$ in the $\mathbf{x}$ direction until it
reaches the point $B=(\mathbf{z_B},\mathbf{z_B})$, this time moving at the
speed of light. The time elapsed is exactly the time needed for a light
signal to travel from $A$ to $B$, but the interaction has been propagated
entirely through extra-space: at intermediate times, the effect of the
interaction is not observable at any physical point between $A$ and $B$.
This seems to be a non-local behaviour with physical effects, and it would
be interesting to investigate if this kind of non-locality is in some way
related to the non-local features emerging from quantum mechanics.
\end{itemize}

\end{document}